\documentclass[iop]{emulateapj}
\slugcomment{Accepted for publication in the EVLA special issue of ApJL}

\usepackage{amsmath}        
\usepackage{amssymb}

\newcommand{\perbeam}{\,beam$^{-1}$}

\shorttitle{Quiescent radio emission from BHXBs}
\shortauthors{Miller-Jones et al.}

\begin{document}

\title{A deep radio survey of hard state and quiescent black hole X-ray binaries}

\author{J.~C.~A. Miller-Jones\altaffilmark{1}}
\affil{International Centre for Radio Astronomy Research - Curtin University, GPO Box U1987, Perth, WA 6845, Australia}
\email{james.miller-jones@curtin.edu.au}

\author{P.~G. Jonker\altaffilmark{2,3}}
\affil{SRON, Netherlands Institute for Space Research, 3584 CA, Utrecht, the Netherlands}

\author{T.~J. Maccarone}
\affil{School of Physics and Astronomy, University of Southampton, Highfield SO17 IBJ, England}

\author{G. Nelemans}
\affil{Department of Astrophysics, IMAPP, Radboud University Nijmegen,
Heyendaalseweg 135, 6525 AJ, Nijmegen, The Netherlands}

\author{D.~E. Calvelo}
\affil{School of Physics and Astronomy, University of Southampton, Highfield SO17 IBJ, England}

\altaffiltext{1}{NRAO Headquarters, 520 Edgemont Road, Charlottesville, VA, 22903, USA}
\altaffiltext{2}{Harvard-Smithsonian Center for Astrophysics, 60 Garden Street, Cambridge, MA 02138, USA}
\altaffiltext{3}{Department of Astrophysics, IMAPP, Radboud University Nijmegen,
Heyendaalseweg 135, 6525 AJ, Nijmegen, The Netherlands}

\begin{abstract}
We have conducted a deep radio survey of a sample of black hole X-ray binaries in the hard and quiescent states, to determine whether any systems were sufficiently bright for astrometric follow-up with high-sensitivity very long baseline interferometric (VLBI) arrays.  The one hard-state system, Swift J1753.5-0127, was detected at a level of 0.5\,mJy\,beam$^{-1}$.  All eleven quiescent systems were not detected.  In the three cases with the highest predicted quiescent radio brightnesses (GRO J0422+32, XTE J1118+480, and GRO J1655-40), the new capabilities of the Expanded Very Large Array were used to reach noise levels as low as 2.6\,$\mu$Jy\,beam$^{-1}$.  None of the three sources were detected, to $3\sigma$ upper limits of 8.3, 7.8, and 14.2\,$\mu$Jy\,beam$^{-1}$, respectively. These observations represent the most stringent constraints to date on quiescent radio emission from black hole X-ray binaries.  The uncertainties in the source distances, quiescent X-ray luminosities at the times of the observations, and in the power-law index of the empirical correlation between radio and X-ray luminosities, make it impossible to determine whether these three sources are significantly less luminous in the radio band than expected.  Thus it is not clear whether that correlation holds all the way down to quiescence for all black hole X-ray binaries.
\end{abstract}


\section{Introduction}
\label{sec:intro}

Black hole X-ray binaries spend the majority of their duty cycles in a low-luminosity ($<10^{33.5}$\,erg\,s$^{-1}$), quiescent state.  Theoretical models to explain the low luminosities and the observed spectra either consider radiatively inefficient accretion flows \citep[e.g.][]{Nar95}, reduce the inner accretion rate to zero \citep{Nar00,Qua00}, invoke winds or outflows \citep{Bla99}, or some combination of these effects.  More recently, jets have been demonstrated to represent a significant component of the power output of accreting black holes at low luminosities \citep{Fen03,Gal05b}, making the coupling between inflow and outflow processes crucial to understanding the nature of low-luminosity accretion.

Simultaneous multi-wavelength observations \citep{Gal03,Cor03} have shown that the radio and X-ray luminosities ($L_{\rm R}$ and $L_{\rm X}$, respectively) of hard state black hole X-ray binaries are correlated via the non-linear relation $L_{\rm R}\propto L_{\rm X}^{0.6}$, implying a fundamental coupling between the accretion and ejection processes in hard state systems, which was later demonstrated to extend unbroken into the quiescent state \citep{Gal06}.  With the inclusion of a mass term, the correlation was further extended to include supermassive black holes \citep{Mer03,Fal04,Gul09}.  However, the universality of this correlation has been called into question.  From an observational perspective, a population of radio-faint X-ray binaries lying well below the correlation was discovered in the relatively bright hard X-ray state \citep[e.g.][]{Gal07}, although \citet{Cor11} found evidence that this population may rejoin the correlation at low luminosities ($\lesssim5\times10^{-5}L_{\rm Edd}$).  More theoretical work also suggested that the correlation should break down below a fraction $10^{-5}$ of the Eddington luminosity $L_{\rm Edd}$, with the power law index steepening as the synchrotron X-ray emission from the jet begins to dominate the Compton-upscattered emission from the accretion flow \citep{Yua05}.  Although two of the most sub-Eddington systems detected in the radio band, Sgr A$^{\ast}$ and A0620-00, both appear to rule out this model, it is possible that the proposed threshold luminosity below which the correlation steepens is either significantly lower than proposed by \citet{Yua05}, or that it is not the same for all sources.  Thus, the true nature of the coupling between inflow and outflow at the lowest luminosities has been difficult to confirm.

The extremely low luminosity of the quiescent state of black hole X-ray binaries has rendered it inaccessible to radio observations for all but the closest and brightest systems, A0620-00 and V404 Cyg, respectively.  A0620-00, at a distance of only $d=1.06\pm0.12$\,kpc \citep{Can10}, is currently the faintest radio-emitting black hole X-ray binary known, with a measured 8.5-GHz radio flux density of $51\pm7$\,$\mu$Jy and a simultaneously-determined 2--10\,keV X-ray luminosity of only $7.1\times10^{30}(d/{\rm 1.2~kpc})^2$\,erg\,s$^{-1}$ \citep{Gal06}. V404 Cyg, at the recently-measured parallax distance of $2.39\pm0.14$\,kpc \citep{Mil09} has a relatively high quiescent luminosity of $4\times10^{32}$\,erg\,s$^{-1}$ \citep{Cor08}.  It is a persistent radio source with a flat radio spectrum \citep{Gal05a} and a flux density of $\sim0.3$\,mJy, although occasional small flares have been detected in the quiescent state, in which the flux density increases by a factor of 3 on timescales of $\sim30$\,min \citep{Mil08}.  The persistent radio emission is believed to arise from a steady, partially self-absorbed, compact jet \citep[e.g.][]{Bla79}.  The persistent, core-dominated nature of this emission makes black hole X-ray binaries ideal astrometric targets when in their hard or quiescent states.  High-precision astrometric observations with very long baseline interferometry (VLBI) would allow us to measure source proper motions, which can provide information about the formation mechanisms of the black holes \citep[e.g][]{Mir01,Mir03}, and in some cases, model-independent distances \citep[e.g.][]{Mil09}.  Prior to embarking on an astrometric campaign, we need to determine which hard state and quiescent systems are bright enough to be detected with the High Sensitivity Array (HSA).

While several black hole X-ray binaries have been monitored during their decay to quiescence following an outburst \citep[e.g.][]{Jon04b,Jon10}, in every case, the sources have faded below the detection threshold before reaching their stable quiescent luminosity.  Using the enhanced sensitivity of the new wideband receiver (CABB; the Compact Array Broadband Backend) on the Australia Telescope Compact Array (ATCA), \citet{Cal10} attempted to detect the two quiescent systems GRO J1655-40 and XTE J1550-564, finding mildly constraining $3\sigma$ upper limits on their quiescent radio emission of 26--27\,$\mu$Jy at 5.5\,GHz and 47\,$\mu$Jy at 9\,GHz.  Deeper radio observations are required to identify astrometric targets and to accurately determine the nature of the correlation between radio and X-ray emission in the quiescent state.

In this paper, we describe a deep radio survey of quiescent black hole X-ray binaries carried out by the Very Large Array (VLA) in order to identify possible targets for an astrometric observing campaign, assuming the existence of more radio-loud quiescent sources similar to A0620-00, V404 Cyg and Sgr A$^{\ast}$.  We also report on higher-sensitivity follow-up observations of three targets using the wide bandwidths now available with the Expanded Very Large Array \citep[EVLA;][]{Per11}.

\section{Observations and data reduction}

\subsection{VLA}

In order to identify possible targets for an astrometric VLBI campaign, we observed eleven quiescent black hole X-ray binaries using the old VLA system under program code AM986.  The integration times were set to provide a robust detection of the minimum flux density for which an astrometric detection with the HSA would be possible, equating to $\sim5$\,h of time on source with the VLA.  The schedules were submitted to the dynamic queue as a series of 1\,h blocks, such that between one and five epochs of observation were taken for each source.  The observation log is shown in Table~\ref{tab:vla_obs}.  The observations were made at a frequency of 8.46\,GHz, in dual circular polarization mode, with 100\,MHz of bandwidth split equally between two intermediate frequency (IF) pairs.  The array was in its CnB and C configurations.

\begin{deluxetable*}{lllcccc}
\tabletypesize{\scriptsize}
\tablecaption{Observation log for the VLA and EVLA observations of our sample of quiescent black hole X-ray binaries.  Observations with 100\,MHz of bandwidth were taken with the old VLA system, whereas those with 2048\,MHz of bandwidth were taken with the new wideband capabilities of the EVLA.\newline$^a$ Phase connection was too poor to allow for calibration.\label{tab:vla_obs}}
\tablewidth{0pt}
\tablehead{
\colhead{Source} & \colhead{Calibrator} & \colhead{Date} & \colhead{MJD} &
\colhead{Bandwidth} & \colhead{Time on source} & \colhead{Image noise} \\
& & & & (MHz) & (min) & ($\mu$Jy\perbeam)}
\startdata
GRO J0422+322 & J0414+3418 & 2009 Jun 21 & $55003.66\pm0.15$ & 100 & 99.0 & 28.7 \\
& & 2009 Jun 22 & $55004.82\pm0.02$ & 100 & 50.2 & -$^a$\\
& & 2009 Jun 25 & $55007.52\pm0.02$ & 100 & 50.3 & 31.7\\
& & 2009 Jun 26 & $55008.76\pm0.02$ & 100 & 50.0 & 14.6\\
& & 2010 Nov 15 & $55515.14\pm0.07$ & 2048 & 148.6 & 2.8 \\
XTE J1118+480 & J1126+4516 & 2009 Jun 02 & $54984.10\pm0.10$ & 100 & 237.7 & 9.6\\
& & 2010 Nov 24 & $55524.53\pm0.06$ & 2048 & 142.5 & 2.6\\
GRO J1655-40 & J1607-3331 & 2010 Dec 13 & $55543.77\pm0.06$ & 2048 & 146.7 & 4.7\\
H1705-250 & J1751-2524 & 2009 Jun 15 & $54997.30\pm0.08$ & 100 & 122.2 & 16.4\\
& & 2009 Jun 16 & $54998.19\pm0.02$ & 100 & 44.0 & 31.0\\
& & 2009 Jun 19 & $55001.26\pm0.02$ & 100 & 48.8 & 27,6\\
& & 2009 Jun 30 & $55012.19\pm0.02$ & 100 & 43.2 & 27.7\\
Swift J1753.5-0127 & J1743-0350 & 2009 Jun 09 & $54991.41\pm0.02$ & 100 & 47.5 & 21.7\\
& & 2009 Jun 10 & $54992.41\pm0.02$ & 100 & 47.2 & 46.6\\
& & 2009 Jun 15 & $54997.40\pm0.02$ & 100 & 46.2 & 23.0\\
XTE J1817-330 & J1820-2528 & 2009 Jul 12 & $55024.24\pm0.02$ & 100 & 48.5 & 32.3\\
& & 2009 Aug 03 & $55046.20\pm0.04$ & 100 & 94.8 & 20.3\\
& & 2009 Aug 08 & $55051.19\pm0.02$ & 100 & 46.3 & 27.6\\
& & 2009 Aug 15 & $55058.15\pm0.02$ & 100 & 48.5 & 28.1\\
XTE J1818-245 & J1820-2528 & 2009 Jul 07 & $55019.19\pm0.02$ & 100 & 50.2 & 25.6\\
& & 2009 Jul 15 & $55027.27\pm0.02$ & 100 & 50.0 & 29.8\\
& & 2009 Jul 23 & $55035.23\pm0.02$ & 100 & 50.0 & 26.5\\
& & 2009 Jul 24 & $55036.14\pm0.02$ & 100 & 65.0 & 21.7\\
& & 2009 Aug 08 & $55051.15\pm0.02$ & 100 & 47.7 & 25.4\\
V4641 Sgr & J1820-2528 & 2009 Jul 01 & $55013.15\pm0.02$ & 100 & 50.2 & 29.7\\
& & 2009 Jul 23 & $55035.27\pm0.02$ & 100 & 48.2 & 28.1\\
& & 2009 Aug 08 & $55051.11\pm0.02$ & 100 & 50.0 & 25.7\\
XTE J1859+226 & J1850+2825 & 2009 Jun 05 & $54987.57\pm0.02$ & 100 & 47.5 & 23.8\\
& & 2009 Jun 06 & $54988.57\pm0.02$ & 100 & 47.2 & 23.6\\
& & 2009 Jun 07 & $54989.56\pm0.02$ & 100 & 47.2 & 23.3\\
& & 2009 Jun 08 & $54990.56\pm0.02$ & 100 & 47.2 & 22.8\\
& & 2009 Jun 10 & $54992.50\pm0.02$ & 100 & 43.2 & 52.6\\
XTE J1908+094 & J1922+1530 & 2009 Jun 10 & $54992.45\pm0.02$ & 100 & 44.0 & 49.6\\
& & 2009 Jun 11 & $54993.45\pm0.02$ & 100 & 47.3 & 25.7\\
& & 2009 Jun 15 & $54997.48\pm0.06$ & 100 & 131.5 & 17.4\\
& & 2009 Jun 18 & $55000.45\pm0.02$ & 100 & 47.0 & 39.9\\
GS 2000+25 & J1931+2243 & 2009 Jun 12 & $54994.53\pm0.04$ & 100 & 86.7 & 15.8 \\
& & 2009 Jun 13 & $54995.57\pm0.02$ & 100 & 44.2 & 22.1\\
& & 2009 Jun 14 & $54996.50\pm0.04$ & 100 & 86.7 & 15.9 \\
XTE J2012+381 & J2015+3710 & 2009 Jun 06 & $54988.61\pm0.02$ & 100 & 44.3 & 21.3\\
& & 2009 Jun 07 & $54989.61\pm0.02$ & 100 & 44.3 & 23.5\\
& & 2009 Jun 09 & $54991.54\pm0.04$ & 100 & 89.0 & 16.2\\
& & 2009 Jun 10 & $54992.54\pm0.02$ & 100 & 44.3 & 40.2\\
\enddata
\end{deluxetable*}

Data were reduced using the 31Dec08 version of the Astronomical Image Processing System ({\sc aips}) software package \citep{Gre03}. Since the data were taken during the transition to the Expanded Very Large Array (EVLA), the array comprised between 20 and 22 upgraded EVLA antennas. Baseline-based calibration was used to account for the mismatch between the bandpasses of the old VLA antennas and the new EVLA antennas.  The flux density scale was set using one of the primary flux calibrators J1331+3030 or J0137+3309, using the coefficients derived by NRAO staff at the VLA in 1999.2.  Phase calibration was performed using the calibrator sources listed in Table~\ref{tab:vla_obs}.  The calibration was applied to the target data, which were imaged, deconvolved, and subjected to self-calibration in cases where the field contained sufficient flux to adequately constrain the solutions.

\subsection{EVLA}

Following the non-detection of all ten quiescent black hole systems during the VLA campaign, we used the nominal source distances and measured quiescent X-ray luminosities of the sources to identify the three targets with the highest quiescent flux densities predicted from the radio/X-ray correlation of \citet{Gal06}; XTE J1118+480, GRO J0422+322 and GRO J1655-40.  These sources were observed under program AM1014 with the EVLA \citep{Per11}, via the Resident Shared Risk Observing (RSRO) scheme.  We observed in the 4--8\,GHz receiver band with an observing bandwidth of 2.048\,GHz.  The observations used two 1024\,MHz basebands centered at 5.38 and 6.80\,GHz, thus providing the most contiguous possible frequency coverage while avoiding radio frequency interference known to exist below 4.5\,GHz and between 5.93 and 6.27\,GHz.  Each baseband was split into eight 128\,MHz sub-bands.  Data were taken in dual circular polarization mode, using 5\,s integrations.  The array was in its C-configuration, sufficiently extended to ensure that confusion did not prevent us reaching the expected thermal noise-limited sensitivity levels.

Data reduction was performed using the Common Astronomy Software Applications (CASA) package \citep{McM07}.  Bandpass and flux density calibration were performed using the primary flux calibrator (J0137+3309 for GRO J0422+32, and J1331+3030 for XTE J1118+480 and GRO J1655-40).  The amplitude scale was set according to the Perley-Butler coefficients derived from recent measurements at the EVLA.  Amplitude and phase gains were derived for all calibrator sources, and the phase calibrators used are shown in Table~\ref{tab:vla_obs}.  Finally, the calibration was applied to the target sources, which, following frequency-averaging by a factor of 8, were then subjected to several rounds of imaging and phase-only self-calibration.  To successfully image and deconvolve such wide bandwidths, we used the CASA implementation of the algorithm for multi-frequency synthesis imaging \citep{Sau94}.

\section{Results}

In our original VLA survey, only Swift J1753.5-0127 was detected, with a flux density of 0.4--0.5\,mJy in each of the observations \citep[the exact flux densities have already been reported by][]{Sol10}.  The other ten sources were not detected, and are therefore too faint for an astrometric HSA observing campaign.  Of the three systems that we observed with the EVLA (GRO J0422+32, XTE J1118+480, and GRO J1655-40), none were detected at a statistically significant level.  The observations of GRO J1655-40 were hampered by the array configuration (C configuration; given the low declination of the source, a hybrid array with an extended northern arm would have been more appropriate) and the presence of significant extended emission in the field.  To filter this out, we selected only data with a {\it uv}-spacing of $>5$\,k$\lambda$.  Our best radio upper limits on all eleven sources in our sample are shown in Table~\ref{tab:final}.

\begin{deluxetable*}{lcccccc}
\tabletypesize{\scriptsize}
\tablecaption{Quiescent luminosities and final $3\sigma$ upper limits on radio emission from our sample of black hole X-ray binaries, after stacking all data on each source.  Distances taken from \citet{Jon04a} except for XTE J1908+094, which is taken from \citet{Jon04b}.  Stellar spectral types are taken from \citet{McC06}.  Quiescent X-ray luminosities are taken from the following references: G01 -- \citet{Gar01}; M03 -- \citet{McC03}; K02 -- \citet{Kon02}; H03 -- \citet{Ham03}; P08 -- \citet{Psz08}; C10 -- \citet{Cal10}; M99 -- \citet{Men99}; T03 -- \citet{Tom03}; J04 -- \citet{Jon04b}.\label{tab:final}}
\tablewidth{0pt}
\tablehead{
\colhead{Source} & \colhead{Distance} & \colhead{Stellar type} & \colhead{$L_{\rm 3-9\,keV}$} & \colhead{$3\sigma$ radio upper} & \colhead{$L_{\rm 8.4\,GHz}$} & \colhead{Reference} \\
& (kpc) & & (erg\,s$^{-1}$) & limit ($\mu$Jy\perbeam) & (erg\,s$^{-1}$) & }
\startdata
GRO J0422+322 & $2.75\pm0.25$ & M2V & $3.2\times10^{30}$ & 8.3 & $<6.3\times10^{26}$ & G01\\
XTE J1118+480 & $1.8\pm0.6$ & K5/M0V & $1.2\times10^{30}$ & 7.8 & $<2.5\times10^{26}$ & M03\\
GRO J1655-40 & $3.2\pm0.2$ & F3/F5IV & 1.0--6.0$\times10^{31}$ & 14.2 & $<1.5\times10^{27}$ & G01,K02,H03,P08,C10\\\
H1705-250 & $8.6\pm2.0$ & K3/7V & $<10^{33}$ & 36.1 & $<2.7\times10^{28}$ & M99\\
XTE J1817-330 & & & & 38.5 & & \\
XTE J1818-245 & & & & 33.5 & & \\
V4641 Sgr & $9.6\pm2.4$ & B9III & $8.0\times10^{31}$ & 47.3 & $<4.4\times10^{28}$& T03\\
XTE J1859+226 & $6.3\pm1.7$ & & $2.6\times10^{30}$ & 38.4 & $<1.5\times10^{28}$ & T03\\
XTE J1908+094 & $\sim8.5$ & & $\sim1.1\times10^{32}$ & 39.8 & $<2.9\times10^{28}$ & J04\\
GS 2000+25 & $2.7\pm0.7$ & K3/K7V & $1.6\times10^{30}$ & 30.7 & $<2.2\times10^{27}$ & G01\\
XTE J2012+381 & & & & 47.4 & \\
\enddata
\end{deluxetable*}

While not formally significant, the measured radio brightnesses at the known source positions of XTE J1118+480 and GRO J1655-40 were $6.4\pm2.6$ and $9.0\pm4.7$\,$\mu$Jy\perbeam, respectively.  This suggests that these are good candidates for deeper observations, using either longer integration times or further increased bandwidth following the completion of the EVLA upgrade.  Considering the entire sample, a Kolmogorov-Smirnov test on all eleven non-detections gave a probability of 0.126 that the measured pixel values at the source positions (as a fraction of the image noise) are consistent with the expected Gaussian distribution of zero mean and unit standard deviation.  Thus we cannot formally exclude the slightly positive pixel values at the positions of XTE J1118+480 and GRO J1655-40 being due to chance.

\begin{figure}
\centering
\includegraphics[width=\columnwidth]{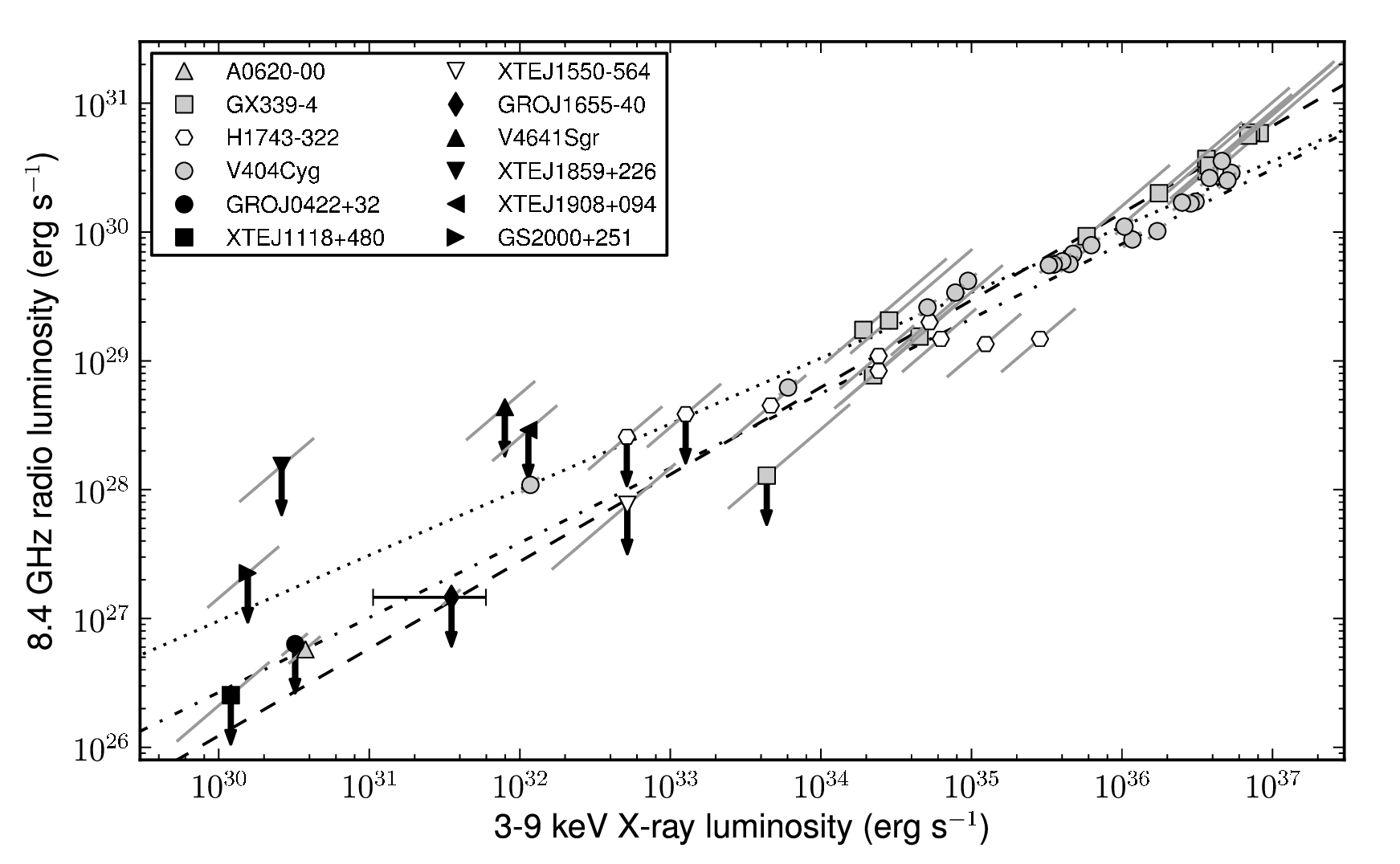}
\caption{Measured radio and X-ray luminosities of black hole X-ray binaries in the hard and quiescent states.  Black markers show the sources studied in this work.  Grey diagonal error bars show the effect of distance uncertainties on the source position in this plane.  The horizontal error bar for GRO J1655-40 shows the measured range of X-ray luminosity for this source.  Dashed line shows the best-fitting correlation ($L_{\rm R}\propto L_{\rm X}^{0.67}$) for the three sources A0620-00, V404 Cyg and GX 339-4 (indicated by grey markers) taken together.  Open markers show other points from the literature for which both simultaneous radio and X-ray observations and distance constraints are available \citep{Cor03,Gal06,Cor08,Cal10,Cor11}.  Dotted line shows the correlation found by \citet{Cor08} for V404 Cyg.  Dot-dashed line shows the correlation fitted by \citet{Gal06}.  The hard-state radio-quiet outliers from the correlation \citep[e.g. Swift J1753.5-0127;][]{Gal07,Sol10} have been omitted for clarity.}
\label{fig:lrlx}
\end{figure}

For all the systems in our sample with constraints on the quiescent X-ray luminosity, we have plotted our radio upper limits on the radio/X-ray correlation in Fig.~\ref{fig:lrlx}, together with a sample of other sources from the literature.  We used WebPimms\footnote{http://heasarc.gsfc.nasa.gov/Tools/w3pimms.html} to convert all X-ray luminosities to a 3--9\,keV range, for compatibility with the observations of V404 Cyg and GX 339-4 that form the bulk of the sample \citep{Cor03,Cor08}.  We calculated the 8.4-GHz radio luminosities by multiplying the measured monochromatic luminosities by a nominal observing frequency of 8.4\,GHz, assuming a flat radio spectrum in all cases.  We performed a fit to the correlation using data from A0620-00, GX 339-4, and V404 Cyg, finding $L_{\rm R}\propto L_{\rm X}^{0.67}$.  All three of our EVLA upper limits fall on or above our best fitting correlation, so we cannot constrain whether the correlation continues to hold at the lowest luminosities.

\section{Discussion}

The only quiescent sources with radio detections are A0620-00 \citep{Gal06} and V404 Cyg \citep{Gal05a}.  The fitted correlation in Fig.~\ref{fig:lrlx} shows that the three sources observed with the EVLA could just have been detected if they were as bright as A0620-00 at a similar X-ray luminosity.  However, when taking into account the known variability in quiescent luminosity and the uncertainties in source distances and the fitted correlation index, we cannot definitively state that the sources are more radio faint in quiescence than expected.

\subsection{Uncertainties}

\subsubsection{Source variability}

Quiescent X-ray binaries are known to be variable across the electromagnetic spectrum, from the X-ray \citep{Cor06} through optical \citep{Can10} and radio bands \citep{Mil08}.  While the lowest-luminosity quiescent system with existing radio and X-ray detections \citep[A0620-00;][]{Gal06} was observed simultaneously in both bands, the new radio observations reported in this paper were not simultaneous with the X-ray observations used in plotting our seven systems on the radio/X-ray luminosity correlation shown in Fig.~\ref{fig:lrlx}.  For GRO J0422+322 and GRO J1655-40, some of the X-ray observations were taken 10 years prior to our EVLA data \citep{Gar01}.  GRO J1655-40 has been particularly well studied in the X-ray band during its quiescent state \citep{Gar01,Kon02,Ham03,Psz08,Cal10}, showing a 3--9\,keV X-ray flux that varied by a factor of more than 5 over the course of 9 years (Fig.~\ref{fig:lrlx}).

\subsubsection{Distance uncertainties}

The distances to X-ray binaries are typically uncertain by 50\% or more \citep{Jon04a}, adding an additional source of uncertainty to the derived X-ray and radio luminosities.  Since the radio/X-ray correlation of \citep{Gal03} is non-linear ($L_{\rm R}\propto L_{\rm X}^{0.7}$), an error in the assumed distance could cause a source to appear artificially bright or faint in the radio band for its X-ray luminosity.  Fig.~\ref{fig:lrlx} shows the effect of the distance uncertainties on the sources forming the radio/X-ray correlation.  Distance estimates and uncertainties come from \citet{Jon04a}, except for GX 339-4 \citep[6--15\,kpc;][]{Hyn04}, V404 Cyg \citep[$2.39\pm0.14$\,kpc;][]{Mil09}, H1743-322 and XTE J1908+094 (assumed Galactic Center distance of 8.5\,kpc with an uncertainty of 2\,kpc for both).  

\subsubsection{Correlation index}

Different authors, by considering different samples, have found different power-law indices for the radio/X-ray correlation.  The original `universal' correlation found by \citet{Gal03} had a correlation index of $0.70\pm0.20$, whereas the addition of the A0620-00 point refined the index to $0.58\pm0.16$ \citep{Gal06}.  \citet{Cor08} found an index of $0.51\pm0.06$ for V404 Cyg taken alone, although such a shallow correlation is ruled out by both the A0620-00 detection and our new data.  However, the lack of low-luminosity radio detections prevents the correlation index being better constrained, with this uncertainty leading to significant variation in the predicted radio brightnesses of quiescent sources.  Furthermore, it is not clear whether the measured relationship between X-ray and radio luminosity should indeed be a single, unbroken power law down to the lowest quiescent luminosities, as has been inferred from the simultaneous radio and X-ray detection of A0620-00 \citep{Gal06}.  Advection-dominated accretion flow models predict a gradual softening of the power-law photon index in the X-ray band as a source decays to quiescence \citep[e.g.][]{Esi97}.  That softening necessarily leads to changes in the fraction of the accretion luminosity arising in a given X-ray band.  Even if the underlying relationship between accretion luminosity and jet emission were indeed a single power-law, these changes would cause a slight deviation from the underlying power law in any empirical measurement of that relationship.  Equally, as discussed in Section~\ref{sec:intro}, synchrotron-dominated jet models predict a steepening of the relation below a critical X-ray luminosity \citep{Yua05}, although such models do not seem to be applicable to either A0620-00 or Sgr A$^{\ast}$.

In summary, the combination of uncertainties in quiescent X-ray luminosity, source distance, and correlation index is sufficient to explain the non-detections of all three of the sources we observed with the EVLA.  As Fig.~\ref{fig:lrlx} shows, when considering these uncertainties, none of the three sources observed with the EVLA fall significantly below the radio/X-ray correlation fitted by \citet{Gal06}, which used the A0620-00 detection to constrain the correlation index down to the lowest luminosities.  Thus our upper limits cannot be used to robustly constrain models for the radio/X-ray coupling in quiescence.

\subsection{Stellar radio emission}

At the sensitivity levels achieved in our EVLA observations, we are reaching the regime where stellar radio emission might be expected to contribute to the radio brightness of the sources.  \citet{Gue02} showed that the peak luminosities of incoherent gyrosynchrotron emission from stellar radio flares can reach $10^{18}$\,erg\,s$^{-1}$\,Hz$^{-1}$ for RS CVn, FK Com and Algol systems.  At the distance of the closest of our systems, XTE J1118+480 (Table~\ref{tab:final}), this corresponds to a radio flux density of 0.25\,mJy\,beam$^{-1}$.  However, our lack of detections suggests that any stellar emission is significantly fainter than this.

The spectral types of the donor stars for our target sources are listed in Table~\ref{tab:final}.  The mass donors for the three sources observed with the EVLA are all late-type main sequence or subgiant stars.  Typical radio luminosities for early/mid K stars are in the range $1\times10^{14}$--$3\times10^{15}$\,erg\,s$^{-1}$\,Hz$^{-1}$, with the higher end of the range corresponding to the most rapid rotators.  Although our X-ray binary targets will be tidally locked (with orbital periods in the range 4.1--62.9\,h), there is little correlation between rotation rate and radio flux density in dwarf stars \citep{Gue02}, so the high rotation speeds should have minimal effect.  We would therefore expect radio flux densities of $<2.5$\,$\mu$Jy\,$(d/{\rm 1\,kpc})$ from the donor stars.  While for astrometric purposes the source of the radio emission is unimportant, stellar emission could begin to confuse the radio/X-ray luminosity correlation should we push deeper on these quiescent systems.

\section{Conclusions}

We have made deep radio observations of the fields surrounding twelve confirmed or candidate black hole X-ray binaries.  The only source detected was Swift J1753.5-0127, at a level of 0.4--0.5\,mJy\,beam$^{-1}$.  The typical $3\sigma$ upper limits on the radio brightnesses of the remaining sources were in the range 30--50\,$\mu$Jy\,beam$^{-1}$.  Three targets were selected for follow-up with the EVLA, using 2\,GHz of bandwidth to enhance the sensitivity.  We found $3\sigma$ upper limits of 7.8, 8.3 and 14.2\,$\mu$Jy\,beam$^{-1}$ for XTE J1118+480, GRO J0422+322 and GRO J1655-40, respectively.  We find that most quiescent systems are beyond the reach of current high-sensitivity VLBI arrays, and cannot therefore be used as astrometric targets.  Uncertainties in the source distances \citep{Jon04a}, in the empirical correlation between radio and X-ray emission, and in the X-ray luminosities at the times of the observations imply that we cannot rule out these sources falling on the radio/X-ray correlation of \citet{Gal03}.  However, while deeper constraints on the quiescent jet emission will be possible on completion of the EVLA, the existing upper limits suggest that it may be difficult to disentangle jet emission from stellar radio emission at such low radio luminosities.

\acknowledgments JCAMJ would like to thank Stephane Corbel for making available his original data on GX339-4.  The National Radio Astronomy Observatory is a facility of the National Science Foundation operated under cooperative agreement by Associated Universities, Inc.  This research has made use of NASA's Astrophysics Data System.

{\it Facilities:} \facility{EVLA}, \facility{VLA}.

\end{document}